\documentclass[11pt,a4paper]{article}
\usepackage{caltn,psfig}

\begin{document}

\docnum{IACHEC Report Series\#1}
\title{Summary of the 2013 IACHEC Meeting}
\author{Catherine E.~Grant$^a$, Matteo Guainazzi$^b$, Lorenzo Natalucci$^c$, \\
Jukka Nevalainen$^d$, Paul P.~Plucinsky$^e$, Andrew Pollock$^b$, Steve Sembay$^f$ \\
on behalf of the IACHEC \\ \\
($^a$Massachusetts Institute of
Technology, \\
$^b$European Space Astronomy Centre of ESA, \\
$^c$Istituto di Astrofisica e Planetologia Spaziali (INAF) \\
$^d$Tartu Observatory \\
$^e$Massachusetts Institute of
Technology, Harvard-Smithsonian Centre for Astrophysics, \\
$^f$Department of Physics
and Astronomy, University of Leicester)
}
\date{\today}                   

\maketitle

{\small
{\bf Abstract} 

We present the main results of the 8th International Astronomical Consortium for High Energy Calibration (IACHEC) meeting, held in Theddingworth,
Leicestershire, between March 25 and 28, 2013. Over 50 scientists
directly involved in the calibration of operational and 
future high-energy missions gathered during 3.5~days to discuss the status of the X-ray payload inter-calibration, as well as possible ways to improve it. 
Sect.~4 of this Report summarises our current understanding of the energy-dependent inter-calibration status.
}

\section{The IACHEC}

The International Astronomical Consortium for High Energy Calibration
(IACHEC)\footnote{{\tt http://web.mit.edu/iachec/}} is a group dedicated to supporting the cross-calibration
environment of high energy astrophysics missions with the ultimate
goal of maximising their scientific return. Its members are drawn
from instrument teams, international and national space agencies and
other scientists with an interest in calibration in this
area. Representatives of over a dozen current and future missions
regularly contribute to the IACHEC activities. Support for the IACHEC
in the form of travel costs for the participating members is
generously provided by the relevant funding agencies.  

IACHEC members cooperate within working groups to define calibration
standards and procedures. The scope of these groups is primarily a
practical one: a set of data and results (eventually published in
refereed journals) will be the outcome of a coordinated and
standardised analysis of reference sources (``high-energy standard
candles''). Past, present and future high-energy missions can use these
results as a calibration reference.

The IACHEC meets yearly to report on the progress of the working
groups and define the next year's activities. The inaugural IACHEC
meeting was held in 2006 on neutral ground in Iceland, but since then
has been hosted by local IACHEC members in Europe, the United States
or Asia.  In 2013 the 8$^{th}$ IACHEC meeting was held in the United
Kingdom at the Hothorpe Hall conference centre in Theddingworth,
Leicestershire. The meeting was hosted by the Leicester University
XMM-Newton EPIC instrument team.

The format of the IACHEC meetings includes both plenary sessions where
instrument calibration status, working group summaries and other
topics of interest are presented and parallel splinter sessions where
working groups meet to discuss results and use the opportunity for
face-to-face data analysis sessions

This Report summarises the main results of
the 8$^{th}$ meeting. It is organised as follows: Sect.~2 describes the main results
discussed by each of the Working Groups. Sect.~3
summarises a discussion good statistical practises in analysing calibration data. Sect.~4 summarises the
cross-calibration status.
For more details, readers are referred to the presentations collected at
the IACHEC meeting web page:
{\tt http://web.mit.edu/iachec/meetings/2013/index.html}.

\section{Working Group reports}

 The scope of the IACHEC Working Groups\footnote{{\tt http://web.mit.edu/iachec/wgs/index.html}}
is to define standard procedures for the calibration of high-energy instrumentation in specific areas, such as:
methodology, definition of calibration targets, and requirements for future missions.

Particular emphasis is addressed to the definition of a set of ``high-energy standard candles'', which can be used as reference for the calibration of past, present and future X-ray
instruments. Each type of standard candle is intended to optimally address a specific parameter sub-space in the energy or timing domain. In this framework, the scope of these working
groups is:

\begin{itemize}

\item to agree on a ``standard spectral model'' for each of the high-energy standard candles
\item to maintain a database of cross-calibration results on the high-energy standard candles
\item to maintain a database of standard candle spectral data, through which any high-energy astronomers will be able to reproduce and gain confidence on the results produced by the
IACHEC Working Groups

\end{itemize}

\subsection{CCD}

The CCD Working Group continued its history of providing a forum for
cross-mission discussion and comparison of CCD-specific modelling and
calibration issues.  At IACHEC 2013, we heard talks about Suzaku/XIS,
Chandra/ACIS, XMM-Newton/EPIC-pn and Swift/XRT on a variety of issues.

To begin with, we heard about the particle background characterisation
efforts on Chandra/ACIS and Suzaku/XIS, specifically, the spatial
structure as a function of energy on Chandra/ACIS, and the time evolution at low
and high-energies on Suzaku.  There was interest in having all
missions/instruments provide standard background models or templates, so
users can more easily include a background model in their spectral fits.

Matteo Guainazzi described the new energy scale calibration for XMM-Newton/EPIC-pn in timing mode that will be released shortly as part of SAS 13.
This calibration includes rate-dependent effects and correction for
X-ray loading seen when the bright source contaminates the offset map.  He
also proposed future coordinated observations of a bright X-ray binary
by multiple missions in their own timing modes to better understand the
pulse height redistribution differences between standard and timing
modes.

Claudio Pagani described the Swift/XRT effort to characterise and map
charge traps both in the flight device and in laboratory irradiation tests.
The Swift/XRT is unique among the IACHEC CCD instruments in that they have
developed deep (10-100 eV) traps since launch.  There was some
speculation as to the cause.  CTI correction software does an excellent
job of mitigating most of the lost performance.

Finally, there was a group discussion of how to manage calibration of
long duration missions when our favourite calibration source (Fe-55) has
a half-life of only 2.73 years.  Astrophysical sources can replace some
of the functionality, as can instrumental features or background lines,
if they interact in the same manner as astrophysical X-rays.  Continual
monitoring of gain evolution, for example, may either need to be less
accurate, or may require more calibration time.  Astro-H will carry a
Modulated X-ray Source which will produce multiple spectral lines with
no radioactive decay.  We eagerly await reports of the on-orbit
performance of this device.

\subsection{Galaxy Clusters}

Since clusters of galaxies are stable over human time-scales, and (the nearest
ones) have high X-ray brightness and hard spectra, they are useful standard
candles for X-ray calibration. Thus, one does not have to observe them
simultaneously for cross-calibration purposes, which allows the usage of
archived (non-simultaneous) data to form large samples. We extended our
previous sample of 11 clusters (Nevalainen et al., 2010) to
$\sim$60 clusters from the HIFLUGCS sample for a comparison of XMM-Newton/EPIC
and Chandra/ACIS instruments. The results of this work are preliminary but seem
to be in line with those published earlier.
 
\begin{figure}[h]
\begin{center}
\psfig{file=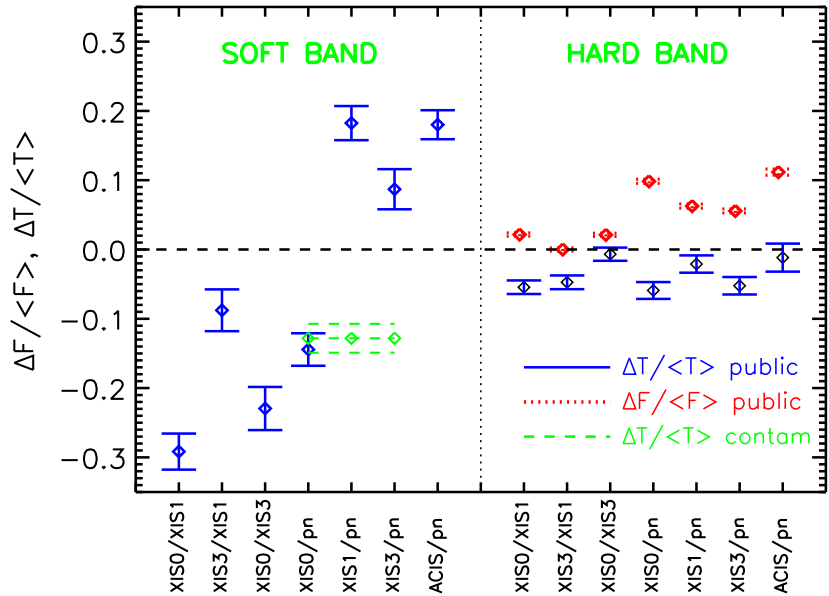,width=12.0cm}
\end{center}
\caption{
Average relative difference ({\it diamonds}), the error of
the mean of the temperatures ({\it solid line}) and fluxes ({\it dotted line})
using the public calibration in the soft band (left side of the plot)
and in the hard band (right side of the plot). Comparison of the
XMM-Newton/EPIC-pn and Suzaku/XIS soft band temperatures using the modification to the
contaminant in the latter is shown with {\it green diamonds} and {\it dashed lines}. Figure after
Kettula et al., 2013
}
\label{fig2}
\end{figure}

We published our results on cross-calibration of the XIS instruments on-board
Suzaku satellite
using clusters of galaxies (Kettula et al. 2013; Fig.~\ref{fig2}). The 2--7~keV
band yields temperatures consistent with EPIC-pn instrument on-board
XMM-Newton satellite. The cluster temperatures in the 2--7 keV band as
measured by XMM-Newton/EPIC-pn have previously been shown to be consistent with those
obtained with XMM-Newton/EPIC-MOS, Chandra/ACIS and BeppoSAX/MECS instruments
(Nevalainen et al. 2010) . Additionally, the continuum-based
bremsstrahlung temperature measurements agree with the ionisation temperatures
obtained from Fe XXV/XXVI line flux ratio measurement performed using
XMM-Newton/EPIC data. Thus, there is a wide consensus on the accurate effective
area calibration among XMM-Newton/EPIC, Chandra/ACIS, Suzaku/XIS and
BeppoSAX/MECS instruments in their hardest energy band.

At the lower energies (0.5--2.0 keV) there are significant cross-calibration discrepancies. We have been
experimenting with using constraints of cluster physics obtained from other
wavelengths, e.g. the thermal pressure via Sunyaev-Zeldovich effect and total
cluster mass via gravitational lensing, to compare with equivalent quantities
derived from X-rays. The aim is to find a consensus between the maximal number
of physical processes and wavelengths to judge, 
which one(s) of the X-ray missions may have larger uncertainties in the
calibration of the effective area. Results are still inconclusive.

We have started a project aiming at comparing the spectroscopic
results for a sample of clusters of galaxies obtained with XMM-Newton/EPIC,
Chandra/ACIS, Suzaku/XIS, Swift/XRT and ROSAT/PSPC instruments. The
preliminary results indicate that this study will be useful in assessing the
uncertainties of the effective area calibration. 

Recently we have started a collaboration with the NuSTAR team in order to
observe some of the clusters already studied in our earlier projects. We are
conducting feasibility studies for the observations which will be useful for
planning new observations.   

\subsection{High-resolution}

The objectives of the High-Resolution Working Group are fourfold: to identify all the lines in the X-ray spectrum first of Capella and later of a small number of reference objects;
feed measured line wavelengths and intensities back to atomic databases; test equilibrium plasma models; and establish X-ray wavelength standards for calibration purposes.
The first spectrum to be addressed is the combined Chandra HETG spectrum of Capella, using both MEG and HEG spectra. The line lists are based on a decomposition of ATOMDB models into
their various line and continuum components, comprising electric and magnetic dipole and quadrupole or EM transitions; dielectronic recombination or DR transitions; two-photon continua
and radiative-recombination continua, all of which involve bounds states of ions; and electron continuum emission. It has been agreed to build models for all 1494~EM lines with a peak
emissivity above 10$^{-19}$~photons~cm$^3$~s$^{-1}$.

\subsection{Non-Thermal SNRs}

The Non-Thermal SNR Working Group is fostering cross-calibration activities using spectra of known ``Non-Thermal'' Supernova Remnants like G21.5-0.9, the Crab, and PSR 1509-58.
A reference paper on G21.5-0.9 was published in the recent past (Tsujimoto et al. 2010) and a cross-calibration project based on X-ray observations of the Crab Nebula is ongoing.  

This year the NuSTAR team has joined the Working Group effort and during the splinter meetings held at this IACHEC Conference a detailed view of the NuSTAR calibration status has been given
(K. Madsen).
Modelling the Crab Nebula spectrum with a simple power-law in the NuSTAR energy band  (4--80~keV), the
residuals are at most within a few percent of the convolved model. This is a good result considering the early mission phase.
In order to achieve a good absolute accuracy an empirical correction of the effective area on
Crab spectra could be necessary in the short term.
A discussion on this issue followed with the basic aim to feed the
NuSTAR team with as much information as possible on the Suzaku and 
INTEGRAL results.
L. Natalucci has presented the status of the Crab cross-calibration project with the detailed list of available observations, with data provided by the Suzaku, RXTE, INTEGRAL and
XMM-Newton teams. In order to cope with the source variability a set of 6 nearly simultaneous epochs (now being updated to 7) have been identified on the basis of the available data-sets. 

The preliminary results on the Crab analysis, mostly based on spectra averaged on a long term period (2005-2011),
confirm the picture that emerged in the Tsujimoto et al. paper, in terms of
normalisation difference between Suzaku/HXD, RXTE/PCA and INTEGRAL/IBIS-ISGRI instruments. More data from INTEGRAL/SPI will be available soon (E. Jourdain)
and possibly from Fermi/GBM (G. Case). Y.~Terada presented recent results on Crab PWN with Suzaku/HXD (Kouzu et al.~2013) indicating spectral variations in the hard X-ray band as well as
X-ray flux. The advantages of global
modelling via a curved power-law instead of a broken power-law were also discussed. The use 
of spectral indices in any given band could be more robust because the slope doesn't change discontinuously. Therefore, comparing fluxes in some pre-defined bandpass for simple 
power-law model
fits could be a valid approach.

Concerning G21.5-0.9, we have pointed out that it makes perfect sense to update the above cited paper with all the additional data from other missions and adding the NuSTAR data, in order
to see if there is evolution and improvements in the cross calibration. L. Natalucci 
presented long-term IBIS/ISGRI spectra showing a remarkable stability
over the years, characterised by $F(15-50$~keV)=($3.95\pm0.24)\times10^{-11}$~erg~cm$^{-2}$~s$^{-1}$ and power law index $\Gamma$=$2.05\pm0.13$. 

Finally, a discussion emerged about using PSR~B1509-58 (the Pulsar in G320.4–01.2, also known as the Hand of God) as a new high energy calibration source. Suzaku, as reported by
Y.~Terada
looked at this source during their first light and is using this as
a timing calibrator (Terada et al. 2008). Possibly the future high energy observatories from China (HMXT) and India (Astrosat) will use it as a calibration target as well.
NuSTAR has plans to observe it as a science target and Suzaku team could be very interested in joining in looking at it. However, it is under discussion
whether PSR~B1509-58 could be used for calibration of NuSTAR due to contamination by the nearby diffuse X-rays.  

For the DAMPE team, J. Li has expressed interest in possible future projects involving higher energy, i.e. gamma-rays in the Fermi band
($\le$300 GeV). An effort for this to get started in
the coming years is envisaged.
       
\subsection{Thermal SNRs}

\begin{figure}[h]
\begin{center}
\psfig{file=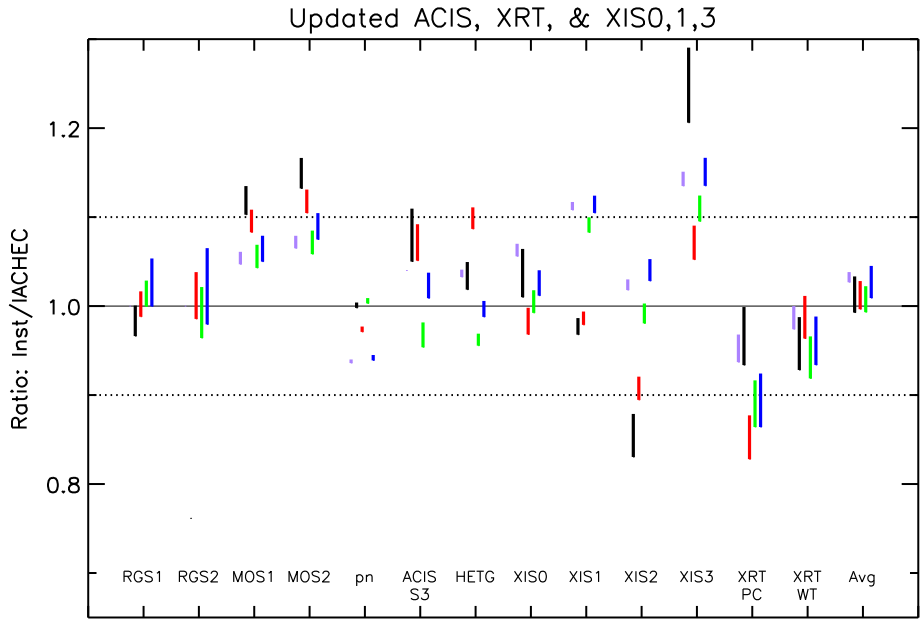,width=12.0cm}
\end{center}
\caption{
Ratio of the O{\small VII}~triplet ({\it black}), O{\small
VIII}~Ly$\alpha$ ({\it red}), Ne{\small IX}~triplet ({\it green}),
and Ne{\small X}~Ly$\alpha$ ({\it blue}) line normalisations, as well as a
global normalisation factor ({\it purple}), when the IACHEC model
is applied to the E0102 spectra of different missions.
}
\label{fig1}
\end{figure}

Work at previous
IACHEC meetings has led to the development of a standard model for the
SMC SNR 1E 0102.2-7219 (E0102 hereafter), which has been described in two SPIE papers
(Plucinsky et al. 2008, 2012). This standard E0102
model has become a valuable tool for the various calibration teams to
understand time-dependent changes in their respective instrument
responses. This model has been used to characterise changes in the
XMM-Newton/EPIC-MOS1 and XMM-Newton/EPIC-MOS2 responses, to monitor the growth of the contamination
layer on the ACIS filter, to develop the contamination model for
the Suzaku/XIS, and to modify the CCD electrode transmission model for the XRT. 
The XMM-Newton/EPIC-pn instrument appears to be the most stable instrument,
based on the consistency of the observed E0102 count rates over the
course of the XMM-Newton mission.  We compared the fitted normalisations
of the O{\small VII}~triplet, the O{\small VIII}~Ly$\alpha$ line, the 
Ne{\small IX}~triplet, and the Ne{\small X}~Ly$\alpha$ line in order to
compare the absolute effective areas of the various instruments at
these energies.  Fig.~\ref{fig1} shows these results relative to the standard
IACHEC model using the latest calibrations or proposed updates to the
calibration, with {\it black} for the O{\small VII}~triplet, {\it red} for the O{\small
VIII}~Ly$\alpha$ line, {\it green} for the Ne{\small IX}~triplet, and the
{\it blue} for the Ne{\small X}~Ly$\alpha$ line (if a global normalisation
for the model spectrum was used, it is shown in {\it purple}).  It should be
emphasised that the IACHEC model is a useful reference model but it
should not be considered the truth since it was developed based on the
XMM-Newton/RGS and Chandra/HETG calibrations at the time of creation.  This comparison
shows that most of the instruments agree to within $\pm10\%$, with a
few exceptions.

The group also continued work on the development of a standard model
for the LMC SNR N132D, the brightest SNR in the LMC. N132D's spectrum
is considerably more complicated than E0102's due to the strong Fe-L
shell emission.  Nevertheless, a prototype standard IACHEC model has
been developed based on the lines identified in the XMM-Newton/RGS and
EPIC-pn spectra.
This
model was compared to the data from the other CCD instruments and some
minor revisions to add weaker lines at energies outside the XMM-Newton/RGS
bandpass will be needed.  The group will work on this revised model
over the coming months.

\section{IACHEC Statistical Methods}

All conclusions regarding values and uncertainties of the physical parameters that encapsulate instruments for calibration purposes and the cosmos at large for more purely scientific
work come from the comparison of data and models. Essentially all high-energy instruments currently and recently in operation accumulate individual photon events and are thus described
by Poisson statistics. For this reason, the IACHEC community recommends and has adopted the C-Statistic rigorous implementation of the Poisson-likelihood for exploration of parameter
spaces rather than any of the alternative Gaussian approximations. A choice needs to be made because of differences in model values inferred with different methodologies. In many cases,
the Poisson-likelihood does not suffer the biases that compromise Gaussian methods.

Analysis typically involves simultaneous use of two observed spectra, both subject to Poisson statistics: one combining source and background; the other from the background only.
C-Statistic analysis provides the most reliable means for constructing models through parameter-space exploration, subject to a variety of constraints. These methods would benefit from
the provision by calibration teams of background models to improve and constrain the implicit phenomenological background methods in regular use. Parameter-space exploration and
goodness-of-fit are independent procedures: the Pearson statistic with model variances provides a reliable goodness-of-fit measure for all Poisson model values greater than zero. In this
overall scheme, rebinning is not relevant and therefore always to be avoided for the loss of information incurred.

Statistical topics have been the subject of regular discussion at IACHEC meetings. This year's 8th meeting enjoyed the benefit of a presentation by David van Dyk, 
Professor of Statistics at Imperial College London, who described Bayesian methods for quantifying effects of calibration uncertainties on scientific analysis. His talk was followed by
round-table discussion exploring a number of topics, some of which were not resolved. An IACHEC Statistics Working Group is to be established without delay to encourage best statistical
practice as a complement to the standard physical models in widespread use in the high-energy calibration community. Details will be found on the IACHEC web pages shortly.

\section{Summary of the cross-calibration status}

In this Section, we summarise the status of inter-calibration among
operational instruments in three energy bands: ``soft'' (photon energy, $E$$\le$2~keV),
``medium'' ($E$$\simeq$2--10~keV), and ``hard'' ($E$$>$10~keV).

\noindent
{\bf Soft}: energy-dependent cross-calibration discrepancies in this energy band were reported by Nevalainen et al. (2010) in the framework
of their study of a sample of bright, nearby Galaxy Clusters observed by Chandra and XMM-Newton. Recent results confirm that the ratio between
the Chandra/ACIS and XMM-Newton/EPIC-pn fluxes increase from -10\% to +10\% going from 0.5 to
2~keV\footnote{cf. {\tt http://web.mit.edu/iachec/meetings/2013/Presentations/Nevalainen.pdf}} . A similar behaviour is observed when
comparing Swift/XRT and XMM-Newton/EPIC-pn. On the other hand, the flux ratio between the Suzaku/XRT and XMM-Newton/EPIC-pn
cameras is energy-independent,
and comprised between -5\% and -10\% (Kettula et al. 2013). XMM-Newton/EPIC-MOS cameras yield fluxes which are on the average in good agreement with
EPIC-pn\footnote{cf. {\tt http://web.mit.edu/iachec/meetings/2013/Presentations/Read.pptx}},
although the exact value in a given observation depends on the calibration of time-dependent EPIC-MOS
redistribution and contamination, which is
still being refined. The accuracy in the description of contaminants in Chandra/ACIS, Suzaku/XIS, and Swift/XRT is also crucial.
Significant
improvements have been recently achieved in these areas. However, more work is required to achieve an overall cross-calibration level better than
10\%, as well as to ensure an accurate description of future evolutions of the contaminants' depth and (possibly) composition.
The comparison of the line normalisation in the multi-mission study of the thermal SNR 1E0102-72, upon which many contamination studies
have been primarily based, already shows an agreement within $\pm10$\% (Plucinsky et al.
2012\footnote{cf. also {\tt http://web.mit.edu/iachec/meetings/2013/Presentations/Thermal\_SNR.pdf}}).

\noindent
{\bf Medium}: Extending the original study by Nevalainen et al. (2010) to a sample of over 60 galaxy clusters in the HIFLUGCS sample confirms an
excellent agreement between the temperatures measured by the Chandra/ACIS and the XMM-Newton/EPIC (within a few
percent over the whole range between $kT$$\simeq$2 to
$\simeq$8~keV\footnote{cf. {\tt http://web.mit.edu/iachec/meetings/2013/Presentations/Schellenberger.pdf}}). Suzaku/XIS cameras yield slightly lower 
temperature (by $\simeq$5\%; Kettula et al., 2013).

\noindent
{\bf Hard}: Recent work on the Crab Nebula confirms the results published in Tsujimoto et al. (2011) on G21.5-0.9.
20--100~keV fluxes measured by Suzaku/HXD, INTEGRAL/IBIS-ISGRI, INTEGRAL/SPI, and RXTE/PCA are
within $\pm$6\%. Swift/BAT yields fluxes $\simeq$20\% lower than INTEGRAL/SPI. Spectral indices are in excellent agreement ($\pm$0.04). Preliminary
NuSTAR calibrations also yields fluxes in agreement within a few percent when compared to Suzaku. A residual difference in spectral shape
(at a level
$\le$0.05\footnote{cf. {\tt http://web.mit.edu/iachec/meetings/2013/Presentations/Walton.pdf}})
is expected to be cured with the next release of NuSTAR effective area files,
to which an empirical correction based on observations of the
Crab Nebula will be applied\footnote{cf. {\tt http://web.mit.edu/iachec/meetings/2013/Presentations/Mardsen.pdf}}.

\section*{References}

\noindent
Kettula K., et al., 2013, A\&A, 552, 47

\noindent
Kouzu et al., 2013, PASJ, in press (arXiv:1303.7109) 

\noindent
Nevalainen J., et al., 2010, A\&A, 523, 22

\noindent
Plucinsky P., et al., 2008, SPIE, 7011, 68

\noindent
Plucinsky P., et al., 2012, SPIE, 8443, 12

\noindent
Terada Y., et al., 2008, PASJ, 60, S25 

\noindent
Tsujimoto M., et al., 2011, A\&A, 525, 25\footnote{see {\tt http://web.mit.edu/iachec/papers/index.html} for a complete list of IACHEC papers}

\end{document}